\documentclass[aip,
reprint,
amsfonts,amsmath,amssymb]{revtex4-1}

\usepackage{graphicx}
\usepackage{bm}

\usepackage{amsmath}

\DeclareMathOperator*{\argmin}{arg\,min}

\usepackage{color}
\usepackage{soul}

\DeclareUnicodeCharacter{2212}{-}

\definecolor{OGreen}{rgb}{0,0.6,0}

\usepackage{listings}

\begin{document}

\title{A molecular perspective on induced charges on a metallic surface}
\author{Giovanni Pireddu}
\affiliation{Sorbonne Universit\'{e}, CNRS, Physico-chimie des \'Electrolytes et Nanosyst\`emes Interfaciaux, PHENIX, F-75005 Paris}
\author{Laura Scalfi}
\affiliation{Sorbonne Universit\'{e}, CNRS, Physico-chimie des \'Electrolytes et Nanosyst\`emes Interfaciaux, PHENIX, F-75005 Paris}
\affiliation{Fachbereich Physik, Freie Universit\"at Berlin, 14195 Berlin, Germany}
\author{Benjamin Rotenberg}
\email{benjamin.rotenberg@sorbonne-universite.fr}
\affiliation{Sorbonne Universit\'{e}, CNRS, Physico-chimie des \'Electrolytes et Nanosyst\`emes Interfaciaux, PHENIX, F-75005 Paris}
\affiliation{R\'eseau sur le Stockage Electrochimique de l'Energie (RS2E), FR CNRS 3459, 80039 Amiens Cedex, France}


\begin{abstract}
Understanding the response of the surface of metallic solids to external electric field sources is crucial to characterize electrode-electrolyte interfaces. Continuum electrostatics offer a simple description of the induced charge density at the electrode surface. However, such a simple description does not take into account features related to the atomic structure of the solid and to the molecular nature of the solvent and of the dissolved ions. In order to illustrate such effects and assess the ability of continuum electrostatics to describe the induced charge distribution, we investigate the behaviour of a gold electrode interacting with sodium or chloride ions fixed at various positions, in vacuum or in water, using all-atom constant-potential classical molecular dynamics simulations. Our analysis highlights important similarities between the two approaches, especially in vacuum conditions and when the ion is sufficiently far from the surface, as well as some limitations of the continuum description, namely neglecting the charges induced by the adsorbed solvent molecules and the screening effect of the solvent when the ion is close to the surface. While the detailed features of the charge distribution are system-specific, we expect some of our generic conclusions on the induced charge density to hold for other ions, solvents and electrode surfaces. Beyond this particular case, the present study also illustrates the relevance of such molecular simulations to serve as a reference for the design of improved implicit solvent models of electrode-electrolyte interfaces.
\end{abstract}

\maketitle


\section{Introduction}

The properties of metal/electrolyte interfaces result from the interplay between, on the one hand, the electronic response of the metal to the charge distribution arising from the solvent molecules and dissolved ions and, on the other hand, the reorganization of the latter in response to the charge distribution at the surface of the metal. From a classical point of view, in perfect metals there is no polarization charge or electric field within the material and the excess or default of electronic density in the presence of an external perturbation is localized only at the surface of the metal. On the electrolyte side, under the effect of thermal fluctuations, the ionic charges are also screened by polar solvents, as well as the ionic cloud formed by the other ions. 

In the continuum picture, the charge density induced at the surface of a perfect metal by a point charge $q_{ion}$ in a semi-infinite medium characterized by a relative permittivity $\epsilon_r$ ($\epsilon_r=1$ for vacuum) is given by~\cite{jackson_classical_1975}
\begin{align}
\label{eq:continuum}
    \sigma_{ind}^{cont}(r) = - \frac{q_{ion}}{2\pi \epsilon_r} \frac{z_{ion}}{(r^2 + z_{ion}^2)^{3/2}} \; ,
\end{align}
with $z_{ion}$ the distance of the ion from the metallic surface and $r$ the radial distance from the ion along the surface. This surface charge distribution corresponds to the electric field at the interface between the dielectric medium and the metal, which is identical, outside of the solid, to the field arising when the metal is replaced by a medium with permittivity $\epsilon_r$ and a so-called ``image charge'' $-q_{ion}$ placed symmetrically to the ion with respect to the interface.
The effect of the solvent is thus limited to decreasing the induced charge by a factor $\epsilon_r$.  Such a description neglects many features of real interfaces, including the atomic and electronic structure of the metallic solid as well as the molecular nature of the solvent and ions of the electrolyte.

Importantly, the presence of an interface modifies the dielectric response of a polar solvent~\cite{ballenegger_local_2003,ballenegger_dielectric_2005} and classical molecular simulations allowed to compute permittivity profiles for interfacial and confined water. This in turn modifies other interfacial properties, such as the capacitance or electrokinetic effects~\cite{bonthuis_dielectric_2011,schlaich_water_2016,loche_breakdown_2018,loche_universal_2020,santos_consistent_2020,motevaselian_universal_2020,jimenez-angeles_nonreciprocal_2020,olivieri_confined_2021}.
In addition, the response of the electronic distribution of the metal to an external perturbation, in particular the fact that its interface with vacuum is not infinitely sharp has been considered within Density Functional Theory (DFT), already in early studies in a simplified 1D geometry~\cite{lang1973a} and nowadays with more advanced functionals and atomically resolved surfaces~\cite{jung_self-consistent_2007,yu_abinitio_2008}. 

In order to investigate the interface between electrodes and electrolytes taking into account atomic and molecular features, it is possible to resort to DFT-based \textit{ab initio} molecular simulations~\cite{sakong_electric_2018,schwarz_electrochemical_2020,sakong_water_2020,le_modeling_2021,le_modeling_2021b}. The computational cost associated with such studies, which accurately describe the electronic density on both the metal and electrolyte sides, renders the sampling of the configurations of the latter difficult, especially for large systems.
Several strategies have been introduced to capture the electronic response of the metal in classical molecular simulations (see Ref.~\citenum{scalfi_molecular_2021} for a recent review). Such methods include descriptions based on fluctuating charges~\cite{siepmann_influence_1995,reed_electrochemical_2007,willard2009a,pastewka2011a,onofrio_voltage_2015,nakano_chemical_2019}, core-shell models~\cite{iori_including_2008,geada_insight_2018}, explicit image charges~\cite{breitsprecher2015a}, induced charges on surfaces~\cite{boda_computing_2004,tyagi_iterative_2010} or Green functions \cite{girotto_simulations_2017}. 

Such simulations allowed in particular to study the effect of the polarization of the metal on the interfacial structure, dynamics and capacitance, the adsorption of ions and biomolecules, or solid-liquid friction, for a variety of electrodes such as gold, platinum or graphite and liquids, from pure water and solutions of simple salts to water-in-salt electrolytes, polyelectrolytes and biomolecules in solution and ionic liquids\cite{merlet2013b,uralcan2016a,li2018b,geada_insight_2018,ntim_role_2020,bagchi_surface_2020,son_image-charge_2021}. They also emphasized the combined role of the atomic structure of the metal and molecular nature of water in the overall hydrophilic/phobic behaviour of the interface~\cite{limmer_hydration_2013,serva_size_2021}.
From a more fundamental point of view, the fluctuations of the charge of the electrode in constant-potential simulations reflect the statistics of the microscopic configurations in the corresponding thermodynamic ensemble. In particular, fluctuation-dissipation relation linking the variance of the charge distribution to the differential capacitance of the system has already been considered in DFT-based ab initio molecular dynamics simulations~\cite{bonnet2012a}, classical Monte Carlo~\cite{kiyohara2007a} and molecular dynamics~\cite{limmer2013a,merlet2014a,haskins_evaluation_2016,scalfi2020a} simulations, and more recently in Brownian Dynamics with an implicit solvent~\cite{cats_capacitance_2021}.

In the present work, we investigate the charge distribution induced on an atomically resolved metallic surface by a single ion in vacuum or in the presence of a molecular solvent. Using classical molecular simulations, we consider the Na$^+$ cation and the Cl$^-$ anion in water, at various distances from a (100) surface of a model gold electrode. This particular example allows us to highlight some expected limitations of the continuum picture and simultaneously illustrates how such simulations can provide a reference for improved implicit-solvent models of these interfaces. After describing the system and methods in Section~\ref{sec:methods}, we present the molecular simulation results for the charge distribution induced by an ion in vacuum and in water in Section~\ref{sec:MDresults}. We then compare these results with the predictions of continuum electrostatics in Section~\ref{sec:continuum}.

\section{Methods}
\label{sec:methods}

\subsection{Simulation details}

\begin{figure}[hbt!]
\centering
  \includegraphics[width=8cm]{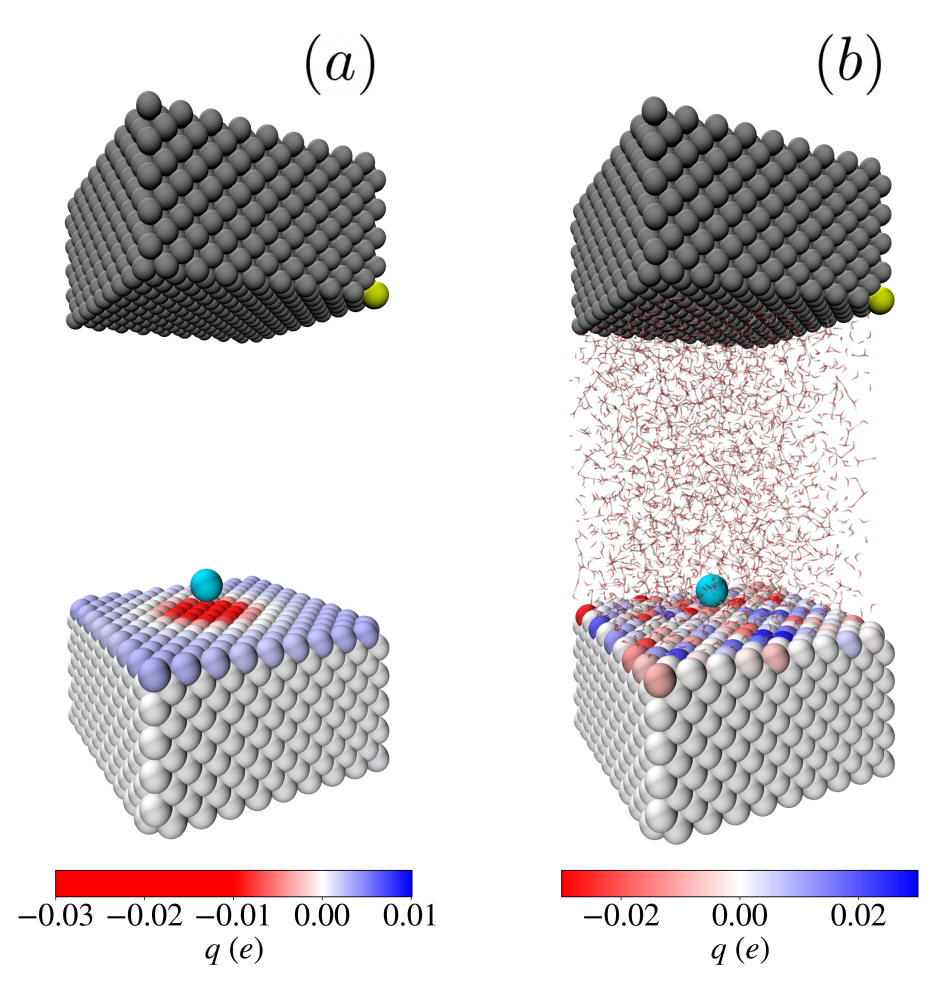}
  \caption{Snapshots of typical molecular configurations considered in this work. (a) The default `vacuum' system. (b) The same including water molecules represented as small sticks. In both panels, the sodium ion, the counterion and the upper wall atoms are depicted as cyan, dark yellow and dark grey spheres respectively. The electrode atoms are colored according to the instantaneous atomic charge, with red and blue corresponding to the negative and positive values, respectively. In both panels, the ion is located at a distance $z_{ion}=5.40$~\AA\ from the first atomic plane of the electrode.
  }
\label{fig:snapshots}
\end{figure}

In order to investigate the charges induced by an external charge in a metallic surface and the effect of the solvent, we considered the two systems illustrated in Fig.~\ref{fig:snapshots}. Each system embeds two confining walls, separated by a distance $L= 48.935$~\AA. Each wall consists of 1620 atoms on an FCC lattice ($9\times 9\times 5$ unit cells with lattice parameter $a=4.07$~\AA\ corresponding to gold), arranged in 10 atomic planes perpendicular to the $z$ direction and facing the inner part of the system with a $(100)$ plane. Only one wall is treated as metallic, using the fluctuating charge model in which each (electrode) atom is equipped with a Gaussian charge of width $w = 0.40$~\AA, with magnitude $q_i$ determined for each configuration of the ion and solvent molecules in order to impose a constraint of constant potential (we considered without loss of generality a value of 0~V) and the constraint of overall electroneutrality~\cite{siepmann_influence_1995,reed_electrochemical_2007,scalfi2020a}. Specifically, each electrode atom $i$ at position ${\bf r}_i$ contributes to the total charge density as:
\begin{align}
\label{eq:rhoi}
    \rho_i({\bf r}) = \frac{q_i}{(2\pi w^2)^{3/2}} e^{-|{\bf r}-{\bf r}_i|^2/2 w^2}
    \, .
\end{align}
The opposite wall has the same structure but the atoms are treated as neutral.

The simulation box has dimensions $L_x=L_y=36.63$~\AA\ and $L_z=  85.565$~\AA\, and periodic boundary conditions (PBC) were applied in the $x$ and $y$ directions only. The ion (Na$^+$ or Cl$^-$) is fixed at a position $(0,0,z_{ion})$, where $z_{ion}$ is expressed relative to the top electrode plane, which sets the reference $z= 0$. This corresponds to a $(a/2,a/2,\cdot)$ site.  Specifically, we consider $z_{ion}=1.50$, $3.14$, $5.40$, $7.03$ and $15.00$~\AA. This set is based on the density profile of Na$^+$ ions in the proximity of the gold electrode surface, reported in Fig.~4a in Ref.~\citenum{scalfi_semiclassical_2020} for a 1M~NaCl aqueous solution using the same description of the gold electrodes, water molecules and ions as in the present work. In particular, the first position corresponds to an ion in direct contact with the surface, the next three to maxima of the density profile (which decrease in intensity as the distance from the surface increases), while the last one represents a typical position in the `bulk' of the solution. In order to enforce the electroneutrality of the charge distribution outside the metal, a counterion is placed on the surface of the opposite wall  at $(L_x/2,L_y/2,L)$. For the simulations in the presence of solvent, illustrated in Fig.~\ref{fig:snapshots}b, the system also includes 2160 water molecules.

The atoms interact via electrostatic interactions, computed using a 2D Ewald summation method taking into account the Gaussian distributions of the electrode atoms~\cite{reed_electrochemical_2007,gingrich_ewald_2010}, and truncated and shifted Lennard Jones (LJ) potentials. Water molecules are modeled with the SPC/E force field~\cite{berendsen1987a} and the LJ parameters for the Na$^+$ and Cl$^-$ ions are taken from Ref.~\citenum{dang_mechanism_1995} and those for the gold atoms from Ref.~\citenum{berg_evaluation_2017}, with the Lorentz–Berthelot mixing rules. All simulations are performed using the molecular dynamics code Metalwalls~\cite{marin-lafleche_metalwalls_2020}, and the charge on the electrode is determined at each time step using the matrix inversion method~\cite{scalfi2020a}. For the ion in vacuum, a single step is sufficient to determine the charge on the wall atoms. For the ions in water, we performed simulations in the $NVT$ ensemble, using a Nos\'e-Hoover chain thermostat~\cite{martyna1992a} with a time constant of 1~ps to enforce a temperature $T=298$~K; these simulations were run for at least 2~ns (after prior equilibration) using a time step of 2~fs. For the analysis detailed below, electrode charges and solvent configurations were sampled every 100 steps (0.2 ps) and every 1000 steps (2 ps), respectively. The uncertainties on the average atomic charges were estimated as the standard error $\hat{\sigma}_{q_i}$ for each electrode atom. These values were used to construct the upper and lower bounds of the induced charge densities (see Section~\ref{sec:analysis}) by simply considering $\langle q_i \rangle \pm \hat{\sigma}_{q_i}$ instead of the average values.

\subsection{Data analysis}
\label{sec:analysis}

In order to analyze the charge induced within the metal, we reconstruct the 3D charge distribution from the atomic contributions (see Eq.~\ref{eq:rhoi}),
\begin{align}
\label{eq:rhoelec}
    \rho_{elec}({\bf r}) &= \sum_{i\in elec} \frac{\left\langle q_i \right\rangle}{(2\pi w^2)^{3/2}} e^{-|{\bf r}-{\bf r}_i|^2/2 w^2}
    \, ,
\end{align}
where the sum runs over all electrode atoms (in the following, we consider sums per atomic plane or over all planes) and $\left\langle \dots \right\rangle$ denotes an average in the canonical ensemble (such an average is not necessary for the ion in vacuum).  Furthermore, in order to visualize the lateral distribution of the induced charge, we integrate over the $z$ direction to obtain the 2D charge density:
\begin{align}
\label{eq:sigmaind2D}
    \sigma_{ind}(x,y) &= \int \rho_{elec}(x,y,z)\  {\rm d}z
    \, .
\end{align}
and also consider the radial average, to obtain:
\begin{align}
\label{eq:sigmaind1D}
    \sigma_{ind}(r) &= \frac{1}{2\pi r} \iint \sigma_{ind}(x,y) \delta\left( r - \sqrt{x^2+y^2} \right) {\rm d}x{\rm d}y
\end{align}
as a function of the radial distance with respect to the ion. Finally, we also analyze the solvation of the ion close to the surface by computing the charge distribution arising from the solvent molecules as
\begin{align}
\label{eq:rhosolv}
    \rho_{solv}(r,z) &= \left\langle \sum_{k\in solv} q_k \delta(r_k-r)\delta(z_k-z)\right\rangle
    \, 
\end{align}
where the sum runs over solvent atoms with partial charge $q_k$ and position expressed in cylindrical coordinates.

\section{Induced charge distribution}
\label{sec:MDresults}

\subsection{Ion in vacuum}
\label{sec:vacuum}

\begin{figure*}[hbt!]
\centering
  \includegraphics[width=15cm]{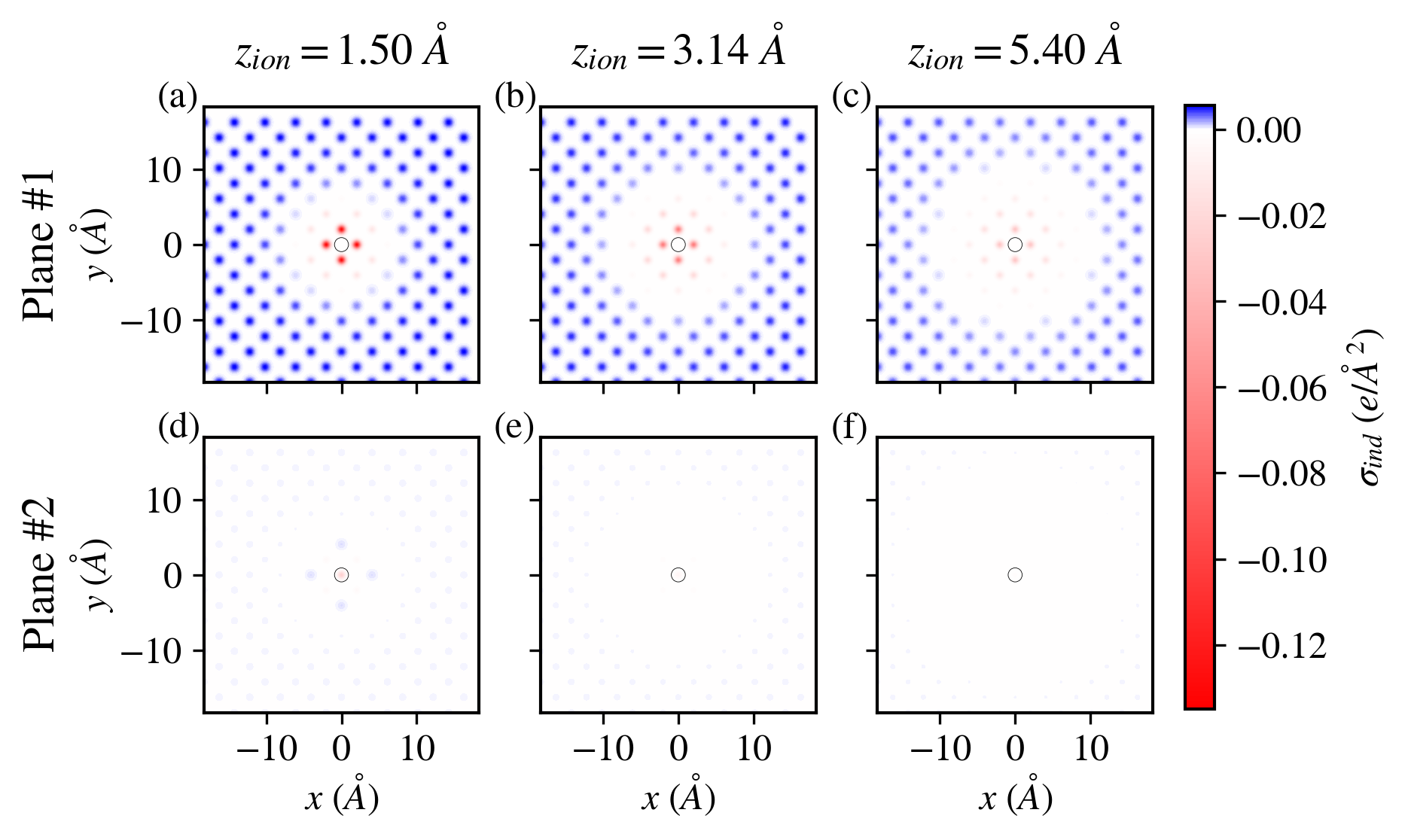}
  \caption{
  Surface charge densities induced on the electrode, $\sigma_{ind}(x,y)$ (see Eq.~\ref{eq:sigmaind2D}) in the first (top, panels a to c) and second (bottom, panels d to f) atomic planes, for a Na$^+$ ion in vacuum. In all panels, the charge density is indicated by the color and the position of the Na$^+$ ion is indicated by a circle.
  }
\label{fig:xymaps_vac_p12}
\end{figure*}

Fig.~\ref{fig:xymaps_vac_p12} shows the 2D distribution, $\sigma_{ind}$ (see Eq.~\ref{eq:sigmaind2D}), of the charge induced in the gold electrode by a Na$^+$ ion in vacuum for three distances $z_{ion}$ of the ion to the top electrode plane. More precisely, panels \ref{fig:xymaps_vac_p12}a, \ref{fig:xymaps_vac_p12}b and \ref{fig:xymaps_vac_p12}c correspond to the charge induced in the first atomic plane, while panels \ref{fig:xymaps_vac_p12}d, \ref{fig:xymaps_vac_p12}e and \ref{fig:xymaps_vac_p12}f correspond to the charge induced in the second atomic plane. The charge induced in the other electrode planes is negligible (see appendix~\ref{sec:zchargedist}), as observed previously~\cite{serva_effect_2021}. In line with the continuum prediction Eq.~\ref{eq:continuum}, in the first plane the induced charge is negative (red color) close to the positive ion and its absolute value gradually decreases laterally away from the ion and as the distance of the ion from the surface increases. However, several features differ from this prediction. 

Firstly, the induced charges becomes positive (blue color) far from the ion, owing to the presence of the counterion and the resulting net electroneutrality of the electrode. Compared to the negative charge induced by the Na$^+$ ion close to the surface, this positive charge induced by the counterion is approximately homogeneous, due to the relatively large distance of this ion from the surface and the PBC in the $x$ and $y$ directions (the effect of PBC will be discussed below). In fact, considering a homogeneous counter-charge distribution instead of a single counterion yielded almost the same results (see appendix~\ref{sec:CAvsUni}). 

Secondly, we observe a square pattern reflecting the atomic lattice. The position of the ion, indicated by the central circle, is above an atom belonging to the second plane, where the magnitude of the induced charge is smaller than in the first, as expected. For the closest ion position (panels \ref{fig:xymaps_vac_p12}a and \ref{fig:xymaps_vac_p12}d), we note the presence of an additional positive charge in the second plane in the vicinity of the ion, probably resulting from the very large negative charge induced on the atoms of the first plane. Such an oscillatory behaviour is reminiscent of Friedel oscillations, even though the description is purely classical in the present case.

\subsection{Ion in water}
\label{sec:solvent}

\begin{figure*}[hbt!]
\centering
  \includegraphics[width=15cm]{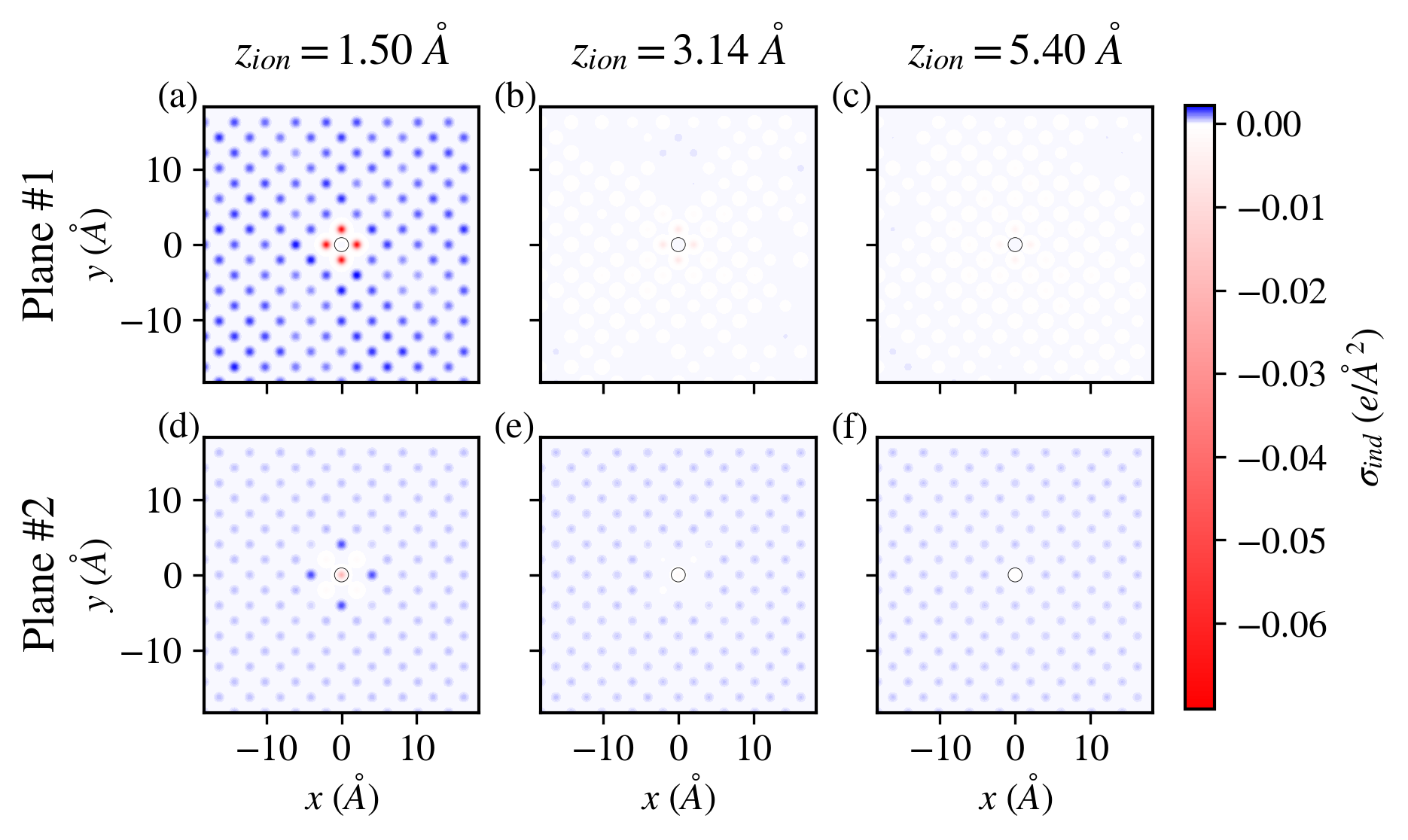}
  \caption{
  Surface charge densities induced on the electrode, $\sigma_{ind}(x,y)$ (see Eq.~\ref{eq:sigmaind2D}) in the first (top, panels a to c) and second (bottom, panels d to f) atomic planes, for a Na$^+$ ion in water. In all panels, the charge density is indicated by the color and the position of the Na$^+$ ion is indicated by a circle. Note that the color scale is different from that of Fig.~\ref{fig:xymaps_vac_p12}.
  }
\label{fig:xymaps_solv_p12}
\end{figure*}

We now turn to the case of a Na$^+$ ion in water, shown for the same ionic positions in Fig.~\ref{fig:xymaps_solv_p12}. Unlike the previous case where a single charge calculation was necessary, the results are now averaged over equilibrium configurations of the solvent. The distribution of the induced charge is qualitatively similar to the vacuum case and the two features related to the presence of the counterion and to the atomic lattice of the electrode are also visible in the presence of the solvent. The main effect is an overall decrease in the magnitude of the induced charge (note the different scale with respect to Fig.~\ref{fig:xymaps_vac_p12}), consistently with the idea of screening of the ionic charge by the polar solvent. However, the extent of this screening depends on the position of the ion with respect to the surface, as discussed below.

\begin{figure}[hbt!]
\centering
  \includegraphics[width=8.5cm]{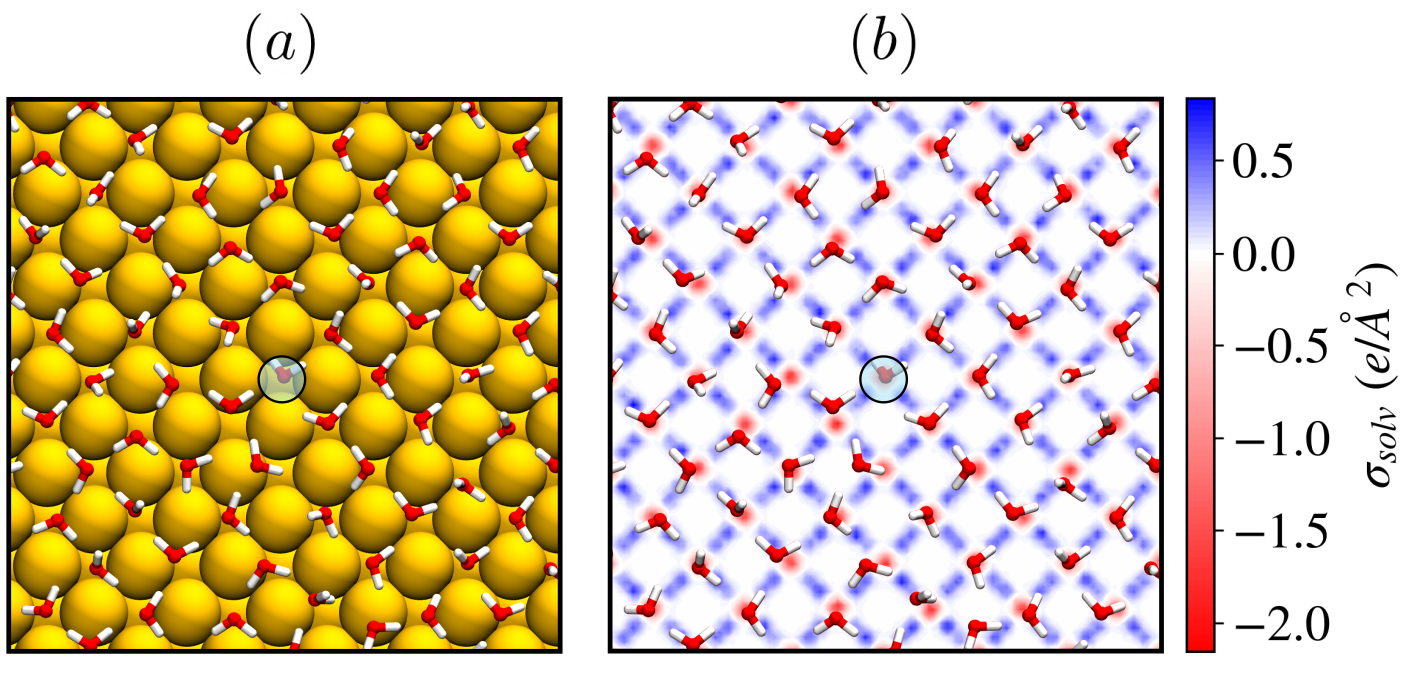}
  \caption{(a) Top view of the water adlayer in a typical configuration; Au atoms are represented in yellow, O atoms in red and H atoms in white, while the position of the Na$^+$ ion (located at a distance $z_{ion}=5.40$~\AA\ from the first atomic plane of the electrode) is shown as a semitransparent cyan circle.
  (b) Same water configuration, shown above the average charge density map arising from water in first the adsorbed layer (see text).
  }
\label{fig:snapshots2}
\end{figure}

Beyond the mere screening of the ionic charge, we also note an important difference with the vacuum case, pertaining to the relative charge density of the first and second planes, most strikingly far from the ion. Indeed, in this region the charge density on the atoms of the second plane is more positive than on the first plane. Such an observation is inconsistent with the continuum picture and arises from the discreteness of water molecules. Indeed, in the first adsorbed layer molecules are located in the cavities formed by four gold atoms in the first plane and one atom in the second plane and form a network of (mainly in-plane) hydrogen bonds as shown for an instantaneous configuration in Fig~\ref{fig:snapshots2}a. The same water configuration is also shown in Fig.~\ref{fig:snapshots2}b together with the average charge density due to molecules in the adlayer (computed from O and H atoms at a distance smaller than 3.6~\AA\ from the first atomic plane, which corresponds to the first minimum of the water density profile). The charge density map indicates the localization of hydrogen atoms between the oxygen basins, \textit{i.e.} a relatively tight H-bond network. This organization of the adsorbed water layer is in turn reflected in the charge induced within the electrode of Fig.~\ref{fig:xymaps_solv_p12}c (for the position of the ion corresponding to the configuration of Fig~\ref{fig:snapshots2}):
While atoms in the first electrode plane are close to both O and H atoms, those of the second electrode layer are closer to the O atoms. The negative partial charge of these O atoms results in a positive charge induced on the atoms of the second plane over the whole surface, except possibly close to the ion. 

\begin{figure}[hbt!]
\centering
  \includegraphics[width=8.5cm]{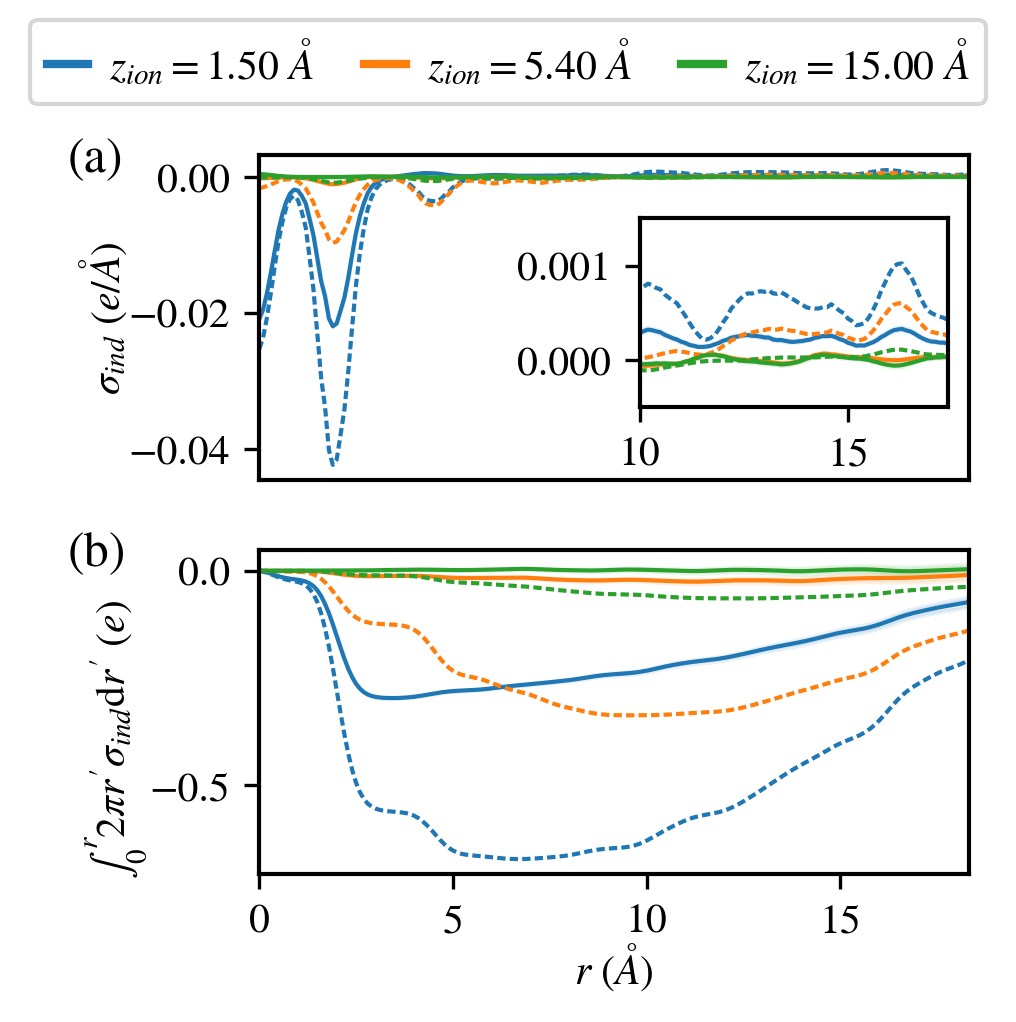}
  \caption{
  (a) Radial charge distribution $\sigma_{ind}(r)$ (see Eq.~\ref{eq:sigmaind1D}) and (b) corresponding radial integral. In both panels, the colors correspond to three distances $z_{ion}$ of the ion with respect to the first atomic plane of the electrode, while dashed and solid lines correspond to a Na$^+$ ion in vacuum and in water, respectively. The shaded areas indicate the standard error (in the aqueous case only) and the inset in panel a is a zoom for large distances.
  }
\label{fig:voidvssolvent}
\end{figure}

The screening effect of the solvent is even clearer when considering the radial charge density profiles, depicted in Fig.~\ref{fig:voidvssolvent}. The results are computed from Eq.~\ref{eq:sigmaind1D} using the total charge distribution, \textit{i.e.} summing over all electrode planes (only the first two planes contribute significantly, see appendix~\ref{sec:zchargedist}), for three distances of the ion from the surface ($z_{ion}=1.5$, $5.4$ and $15.0$~\AA). The oscillations of $\sigma_{ind}(r)$ in panel~\ref{fig:voidvssolvent}a reflect the position of the electrode atoms: the negative minimum at $r=0$ corresponds to the Gaussian centered on the atom below the ion in the second plane, while the next minimum correspond to the nearest neighbours in the first plane. The comparison between the vacuum (dashed lines) and water (solid lines) cases shows the reduction of the induced charge density in the presence of the solvent, which is not identical for all the positions of the ion. In particular, for the closest position, $z_{ion}=1.5$~\AA, there is no water between the ion and the nearest electrode atoms, so that the charge distribution near the atom below in the second plane is hardly changed and the reduction for the nearest neighbours in the first plane is only reduced by a factor of $\approx2$. The decrease in $\sigma_{ind}(r)$ with respect to the vacuum case is more pronounced when the ion is farther from the surface. The role of the water structure will be further examined in Section~\ref{sec:ioniccharge}. 

Fig.~\ref{fig:voidvssolvent}b further illustrates the integrated charge density, which provides a convenient way to analyze the behaviour far from the ion (shown in the inset of Fig.~\ref{fig:voidvssolvent}a). This representation clearly shows the radial distance $r$ at which the radially averaged density $\sigma_{ind}(r)$ changes sign (this corresponds to the minimum of the integrated charge density). This distance systematically shifts towards larger $r$ values when the ion is farther from the surface, consistently with the increase in the spread of the charge distribution induced by the ion (see also Figs.~\ref{fig:xymaps_vac_p12} and~\ref{fig:xymaps_solv_p12}).
The integrated charge densities also better show that the overall screening effect is less pronounced when the ion is very close to the surface. In order to go beyond such a qualitative statement, we now compare these molecular simulation results with the predictions of continuum electrostatics.

\section{Comparison with continuum electrostatics}
\label{sec:continuum}

\subsection{Ion in vacuum}

\begin{figure}[ht!]
\centering
  \includegraphics[width=8.5cm]{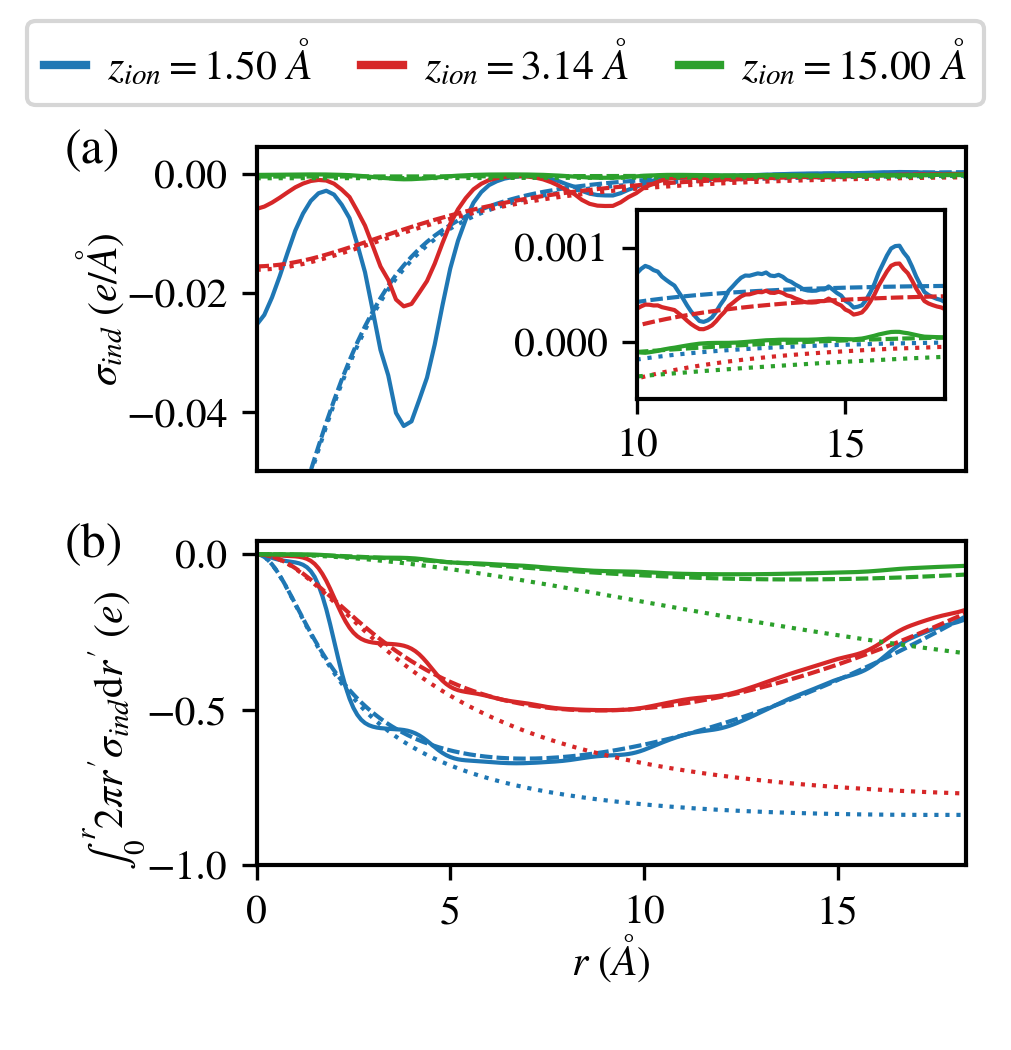}
  \caption{
    (a) Radial charge distribution $\sigma_{ind}(r)$ (see Eq.~\ref{eq:sigmaind1D}) and (b) corresponding radial integral, for a Na$^+$ ion in vacuum. In both panels, the colors correspond to three distances $z_{ion}$ of the ion with respect to the first atomic plane of the electrode. Solid lines correspond to the simulation result, while dashed (resp. dotted) lines correspond to the continuum predictions Eq.~\ref{eq:continuum} taking (resp. not taking) into account the periodic images. The inset in panel a is a zoom for large distances. }
\label{fig:voidvsan}
\end{figure}

Fig.~\ref{fig:voidvsan} compares the atomistic results (solid lines) with the continuum prediction, Eq.~\ref{eq:continuum}, for a Na$^+$ ion in vacuum ($\epsilon_r=1$) and three distances $z_{ion}$ from the top electrode plane. The direct application of Eq.~\ref{eq:continuum}, \textit{i.e.} considering only the ion in the simulation box and the corresponding counterion (dotted lines), fails to reproduce the oscillations of the induced charge density profiles, as described above, but its integral coincides reasonably well with the numerical results at short distance. However, it significantly deviates beyond $r\approx5$~\AA. This observation is not due to the precise location of the counterion (which is much further from the surface than the ion of interest), as we obtain almost identical results placing it in one corner of the simulation box, aligned with the ion in the center of the box, or replacing it by four charges $-q_{ion}/4$ at the corners or the box (not shown).

Rather, the discrepancy is essentially due to the fact that the simulated system is periodic in the $x$ and $y$ directions, so that the ionic distribution corresponds in fact to a 2D-periodic array of ions and counterions. The necessity to take periodic images into account to compare with the analytical prediction was already pointed out by Reed \textit{et al.}~\cite{reed_electrochemical_2007}. We do this numerically by considering 41$\times$41 ion/counterions, \textit{i.e.} the central ones and 20 images in all directions, which is sufficient to converge the sum. The prediction of Eq.~\ref{eq:continuum} for the integrated charged density (panel~\ref{fig:voidvsan}b) is then in good agreement with the atomistic results over the whole range of considered distances, except for the oscillations at short distance due to the atomic structure of the electrode, as expected. In the following, we therefore compare with the continuum prediction in the presence of water only taking into account the periodic boundary conditions.

\subsection{Ion in water}

\begin{figure}[hbt!]
\centering
  \includegraphics[width=8.5cm]{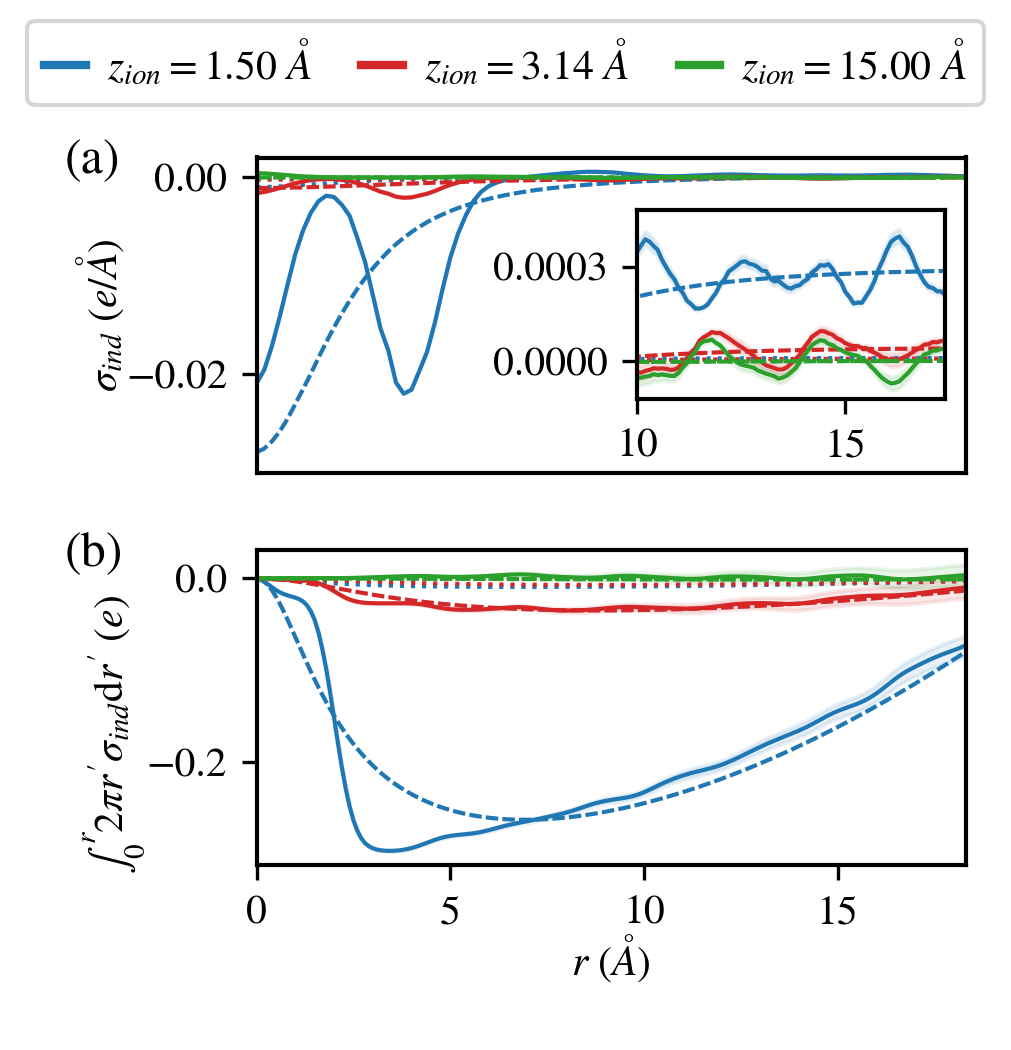}
  \caption{
  (a) Radial charge distribution $\sigma_{ind}(r)$ (see Eq.~\ref{eq:sigmaind1D}) and (b) corresponding radial integral, for a Na$^+$ ion in water. In both panels, the colors correspond to three distances $z_{ion}$ of the ion with respect to the first atomic plane of the electrode. Solid lines correspond to the simulation result, while dotted and dashed lines correspond to the continuum predictions Eq.~\ref{eq:continuum}, taking into account the periodic images, using the bulk permittivity $\epsilon_{r}^{bulk}=70.7$ or a fitted value for each distance (see text), respectively. The shaded areas indicate the standard error (for the molecular simulation results only) and the inset in panel a is a zoom for large distances.
  }
\label{fig:solvvsan}
\end{figure}

We now turn to the case of a Na$^+$ ion in water, illustrated in Fig.~\ref{fig:solvvsan} where the molecular simulation results (solid lines) are compared to Eq.~\ref{eq:continuum} (with the above-mentioned sum over periodic images) using the relative permittivity of the SPC/E water model\cite{ramireddy_dielectric_1989}, namely $\epsilon_r^{bulk}=70.7$ (dotted lines). These predictions are actually difficult to see on the same scale as the MD results, indicating that they excessively underestimate the charge density -- or equivalently overestimate the screening of the charge by the solvent.

Such a poor prediction is not unexpected, since it is now well established that the dielectric properties of water (and other polar liquids) are drastically modified at interfaces. The response of the polarization to an applied field becomes in principle non-local and tensorial, but it is possible to introduce a local permittivity tensor, whose components parallel and perpendicular to the walls are related to the equilibrium fluctuations of the polarization~\cite{ballenegger_local_2003}. These fluctuations can be sampled in molecular simulations to determine how these components depend on the position with respect to the interface and/or on the width of the fluid slab, for confined fluids)~\cite{ballenegger_dielectric_2005,bonthuis_dielectric_2011,bonthuis2012a,schlaich_water_2016,loche_universal_2020,motevaselian_universal_2020,deisenbeck_dielectric_2021}. The general picture emerging from these studies is a reduction of the permittivity in the vicinity of solid walls, consistently with experimental observations on confined water~\cite{fumagalli_anomalously_2018}, even though the microscopic origin of this reduction is not to be found in the molecular structure of interfacial water but rather in the frustration of collective long-range fluctuations~\cite{olivieri_confined_2021}. 

The implications of a decrease in permittivity on other interfacial properties such as the capacitance or the electrokinetic response has also been investigated, in particular in the framework of suitably parameterized slab models~\cite{bonthuis_dielectric_2011,bonthuis2012a,schlaich_water_2016,loche_universal_2020}. In the present case of the charge induced by a single ion, the translational invariance along the surfaces is broken and a continuum description should in principle involve a spatial dependence of the permittivity tensor. As a first step toward a simpler implicit-solvent description, we follow instead an effective approach by considering the relative permittivity $\epsilon_r$ in Eq.~\ref{eq:continuum} as a fitting parameter (see appendix~\ref{sec:relperm}) for each distance $z_{ion}$ of the ion from the surface. The radial density profiles resulting from this procedure (dashed lines) are compared with the MD simulation results in Fig.~\ref{fig:solvvsan}. The agreement is now comparable with that observed in the vacuum case, \textit{i.e.} going through the oscillations due to the atomic structure of the electrode (especially at short distance) and following closely the integrated charge density at larger distances. The values resulting from this fitting procedure are $\epsilon_r^{eff}=2.5$, 14.4, 15.1, 21.4 and 68.0, for $z_{ion}=1.50, 3.14, 5.40, 7.03$ and $15.00$~\AA, respectively. Such a decrease in the effective permittivity as the ion approaches the surface reflects a reduced screening of the ionic charge by the solvent. We emphasize however that, even though this crude approximation seems sufficient to account empirically for the numerical observations, a physically better motivated description of the dielectric response is of course desirable. 

\subsection{Cation vs anion}
\label{sec:ioniccharge}

\begin{figure}[hbt!]
\centering
  \includegraphics[width=8.5cm]{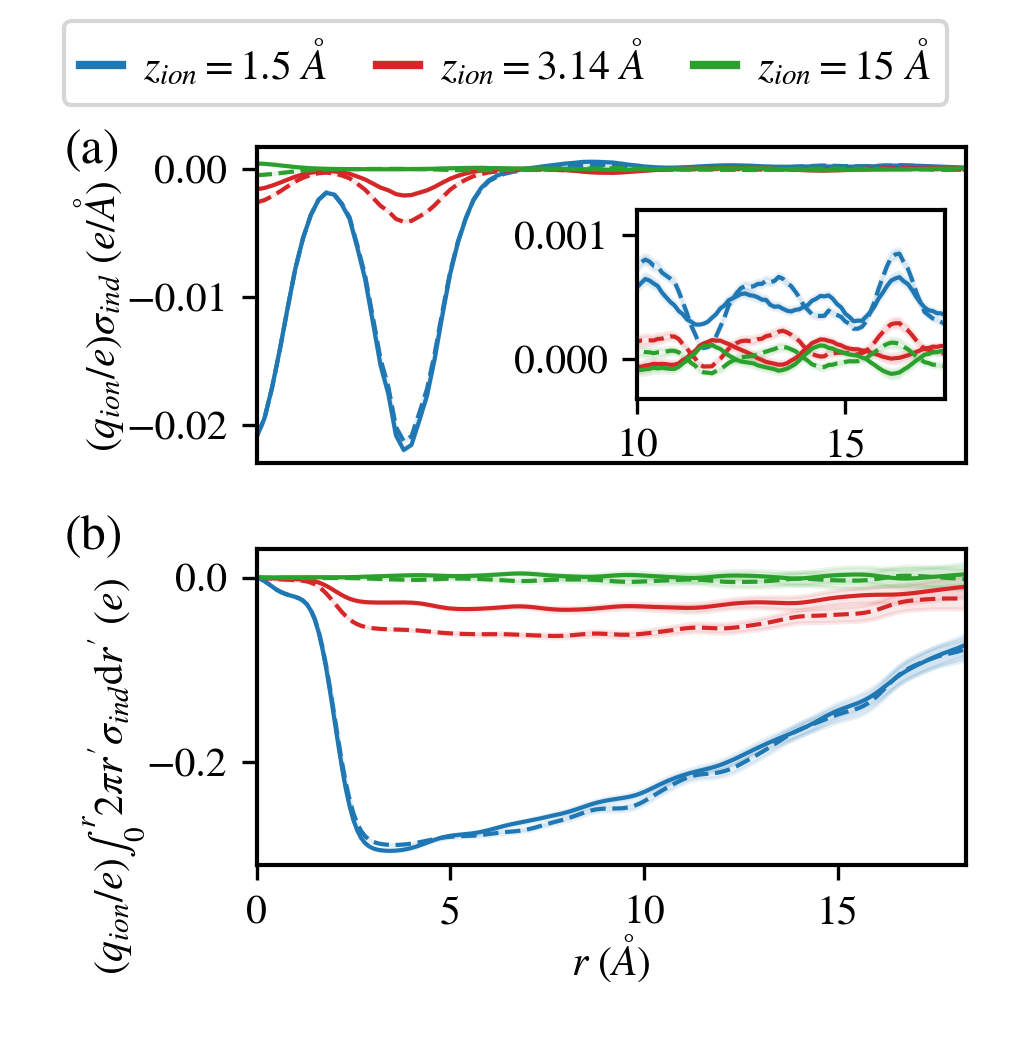}
  \caption{
    (a) Radial charge distribution $\sigma_{ind}(r)$ (see Eq.~\ref{eq:sigmaind1D}) multiplied by the valency of the ion, $q_{ion}/e$, and (b) corresponding radial integral, for a Na$^+$ (solid lines) or Cl$^-$ (dashed lines) ion in water. In both panels, the colors correspond to three distances $z_{ion}$ of the ion with respect to the first atomic plane of the electrode. The shaded areas indicate the standard error and the inset in panel a is a zoom for large distances.
    }
\label{fig:navscl}
\end{figure}

Another important prediction of Eq.~\ref{eq:continuum} is that the charge induced on the metallic surface should be opposite when the sign of the ionic charge is changed. This is indeed the case with an atomically resolved constant-potential electrode when the ion is in vacuum (not shown). In order to compare the molecular simulation results obtained with a Na$^+$ cation or a Cl$^-$ anion in water, the radial charge distribution $\sigma_{ind}(r)$ is shown multiplied by the valency of the ion, $q_{ion}/e$, in Fig.~\ref{fig:navscl} for three ion distances from the first electrode plane. The results for the closest and farthest ion positions ($z_{ion}=1.5$ and $15.0$~\AA, respectively) are consistent with the predicted charge inversion over the whole range of radial distance $r$, despite the differences between the MD and continuum results discussed above. In contrast, the radial charge distribution induced by Na$^+$ or Cl$^-$ at the intermediate distance $z_{ion}=3.14$~\AA\ (red lines) differ by more than a sign inversion. 

\begin{figure*}[hbt!]
\centering
  \includegraphics[width=18.5cm]{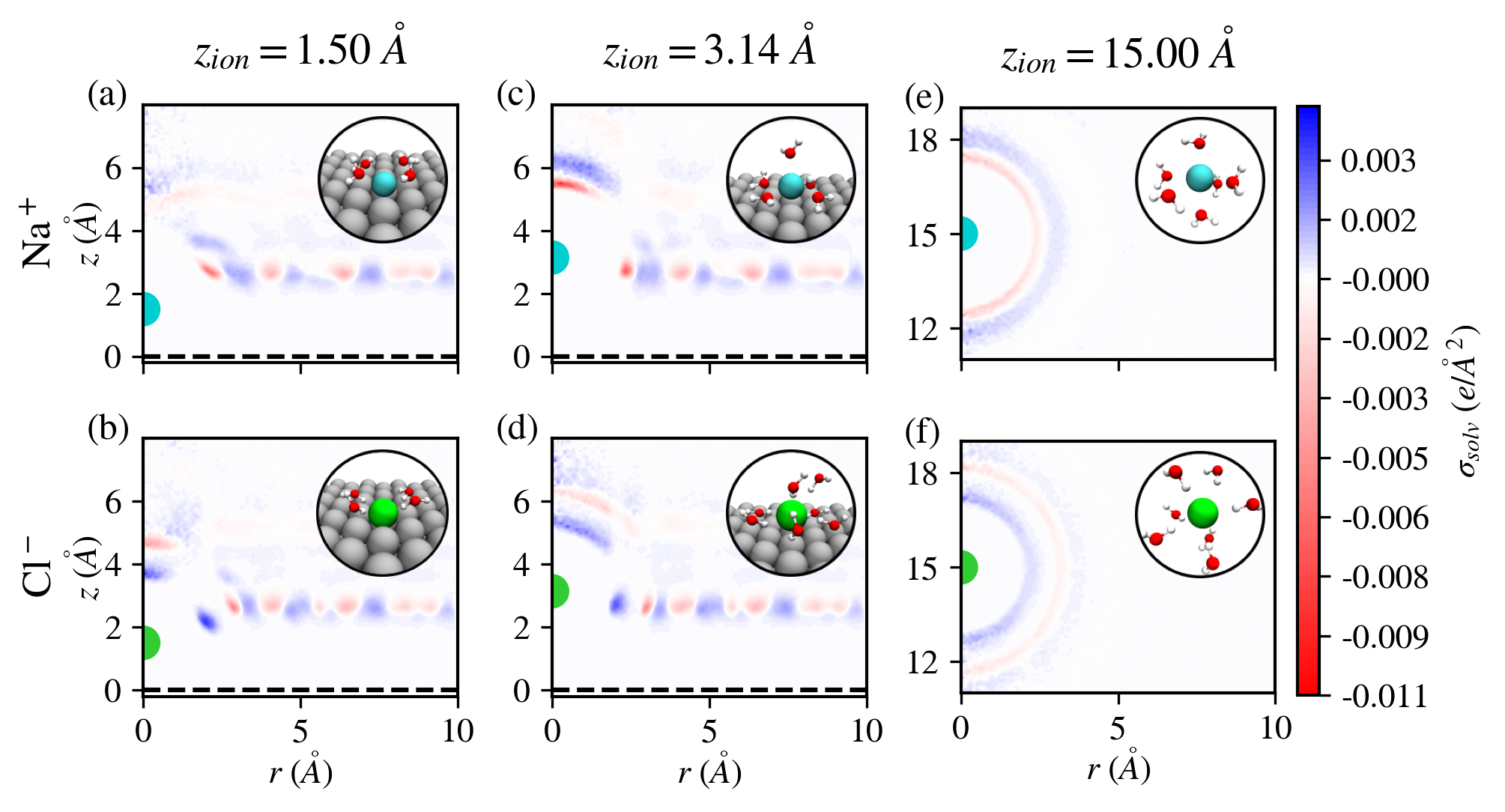}
  \caption{
  Average charge density due to O and H atoms from water molecules in the $(r,z)$ plane, for Na$^+$ (panels a, c and e) and Cl$^-$ (panels b, d and f) ions at three distances from the first atomic plane of the electrode. The color indicates the charge density, with negative values in red and positive value in blue. In each panel, the inset shows a typical configuration of water molecules in the first solvation shell.
    }
\label{fig:rzmapAll}
\end{figure*}

Such a difference originates from the well known asymmetric solvation of ions by water, since water molecules are not simple point dipoles~\cite{grossfield_2005,mukhopadhyay_2012}.
This asymmetry is clearly visible in Fig.~\ref{fig:rzmapAll}, which shows the average charge density due to O and H atoms from water molecules in the $(r,z)$ plane, for Na$^+$ and Cl$^-$ ions at three distances from the first atomic plane of the electrode. For ions far from the surface (panels \ref{fig:rzmapAll}e and \ref{fig:rzmapAll}f for Na$^+$ and Cl$^-$, respectively), the charge density is spherically symmetric around the ion, but the alternating signs of the charge density shells reflect the different typical orientation of water molecules around the ions, with the O atom closer to the Na$^+$ cation and a H-bond donated by water to the Cl$^-$ anion. However, this asymmetry manifests itself only in the close vicinity of the ion, so that the difference between Na$^+$ and Cl$^-$ in Fig.~\ref{fig:navscl} for $z_{ion}=15$~\AA\ is the mere change of sign of the induced charge density. 

The H-bond pattern of water in the first adsorbed layer (see Fig.~\ref{fig:snapshots2}) is also visible in the $(r,z)$ plane, for molecules sufficiently far from the ion, as the alternating positive and negative charge density basins in panels \ref{fig:rzmapAll}a to \ref{fig:rzmapAll}d (the same structure is also present near the surface when the ion is far from the surface, but the corresponding range of $z$ is not visible in panels \ref{fig:rzmapAll}e and \ref{fig:rzmapAll}f). These panels further show that the ion solvation shell interferes with the network of surface water as the ion approaches the surface. This interplay explains why the charge induced by a Cl$^-$ anion is not simply the opposite of that induced by a Na$^+$ cation, at least for the intermediate distance $z_{ion}=3.14$~\AA\ of panels \ref{fig:rzmapAll}c and \ref{fig:rzmapAll}d. Even though such an asymmetry is still present for $z_{ion}=1.50$~\AA\ in panels \ref{fig:rzmapAll}a and \ref{fig:rzmapAll}b, at such a short distance the ion is closer from the surface than the water molecules in its first solvation shell. The charge induced by the ``bare'' ion then dominates the polarization of the electrode, so that the leading effect is again a simple charge inversion when changing the sign of the ionic charge.

\section{Conclusion}

Using molecular simulations, we investigated the charges induced on an atomically resolved metallic surface by a single ion in vacuum or in the presence of a molecular solvent. Specifically, we considered the Na$^+$ cation and the Cl$^-$ anion in water, at various distances from a (100) surface of a model gold electrode. The charge distribution within the electrode qualitatively follows the predictions of continuum electrostatics, in particular (i) the induced charge density decreases with the radial distance from the ion or with an increase of the distance of the ion from the surface; (ii) the induced charged density decreases in the presence of the polar solvent, consistently with the idea of screening by the latter; (iii) the main difference between the cation and anion cases is the reversal of the induced charge density. 

However, the present molecular simulation study also highlights several expected limitations of the continuum picture: (i) the induced charge density oscillates following the atomic lattice of the electrode, even in the absence of solvent; (ii) the induced charge density reflects the structure of the first adsorbed water layer, even far from the ion; (iii) the bulk permittivity of the solvent is not sufficient to capture the screening of the ionic charge as the ion approaches the surface; (iv) the asymmetry in the solvation shell of cations and anions results in effects beyond the mere sign reversal of the induced charge when the ion approaches the interface (even though, as it approaches even further, this leading effect is recovered due to partial desolvation). 

The detailed features of the charge distribution are system-specific (even for a given metal and solvent, the interfacial liquid structure crucially depends on the considered crystal face, see \textit{e.g.} Ref.~\citenum{limmer_hydration_2013} for water on Pt) and depend to some extent on the details of how the metallic character is described in these classical constant-potential simulations (in particular, on the width $w$ of Gaussian charge distribution on atoms~\cite{serva_effect_2021} or a screening length inside the metal~\cite{scalfi_semiclassical_2020}). Nevertheless, we expect the above generic conclusions on the induced charge density to hold for other ions, solvents and electrode surfaces than the ones considered here. In addition, we have considered here a single metallic surface and another natural step is to investigate the effect of voltage between two electrodes on the induced charge distribution, since it will also modify the organization of the interface. Finally, such molecular simulations studies provide a useful reference to design improved implicit solvent models, beyond the bulk (or \textit{ad hoc} distance-dependent) permittivity. First steps in this direction could for example include continuum electrostatics for an interfacial slab model~\cite{loche_breakdown_2018}, molecular density functional theory~\cite{ding_efficient_2017,jeanmairet2019a,jeanmairet2019b} or field theories~\cite{berthoumieux_gaussian_2018,berthoumieux_dielectric_2019,vatin_electrostatic_2021} based on molecular models. 

\section*{Acknowledgements}

The authors are grateful to Mathieu Salanne for many discussions on this and related topics. This project has received funding from the European Research Council  under the European Union's Horizon 2020 research and innovation programme (grant agreement No. 863473). This work was supported by the French National Research Agency (Labex STORE-EX, Grant ANR-10-LABX-0076). 


\section*{Data availability}
The data that support the findings of this study are available from the corresponding author upon reasonable request.


\appendix

\section{Charge distribution across the electrode planes}\label{sec:zchargedist}

\begin{figure}[hbt!]
\centering
  \includegraphics[width=8.5cm]{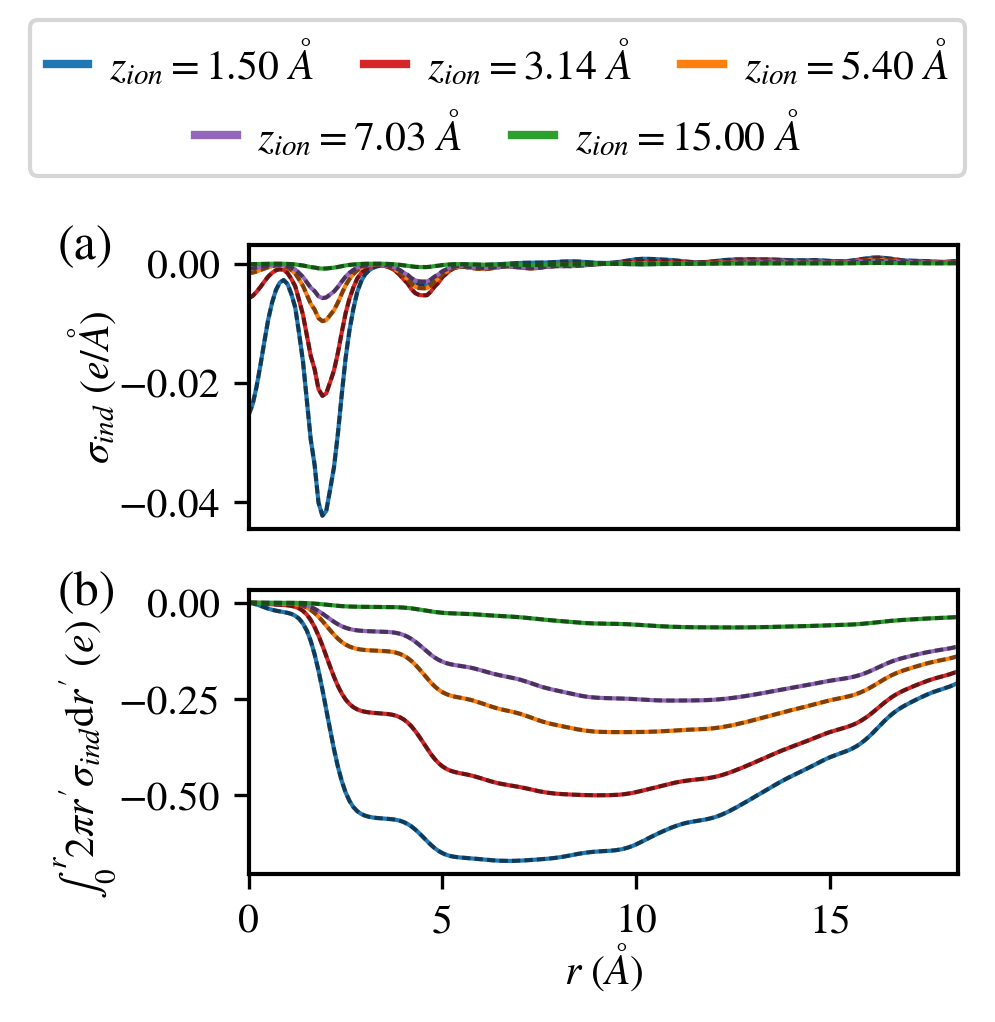}
  \caption{
  (a) Radial charge density profile, $\sigma_{ind}(r)$, considering either the first two electrode planes or the whole electrode, for a Na$^+$ ion at a distance $z_{ion}=1.50$~\AA\ from the first atomic plane. (b) Radial integral of $\sigma_{ind}(r)$. In both panels, dashed (resp. solid) lines correspond to the ion in vacuum (resp. in water) and results for the first two planes and for the whole electrode are indicated with blue lines and orange lines, respectively.
  }
\label{fig:allvs12}
\end{figure}

As discussed in the main text, most of the induced charge is localized in the 
first two electrode planes. This can be easily observed by comparing the radial charge density profiles, $\sigma_{ind}(r)$, when taking into account either only the first two atomic planes or the whole electrode, as shown in Fig.~\ref{fig:allvs12}. The agreement between the two data sets is almost perfect.

\section{Influence of the distribution of the counter-charge}\label{sec:CAvsUni}

\begin{figure}[hbt!]
\centering
  \includegraphics[width=8.5cm]{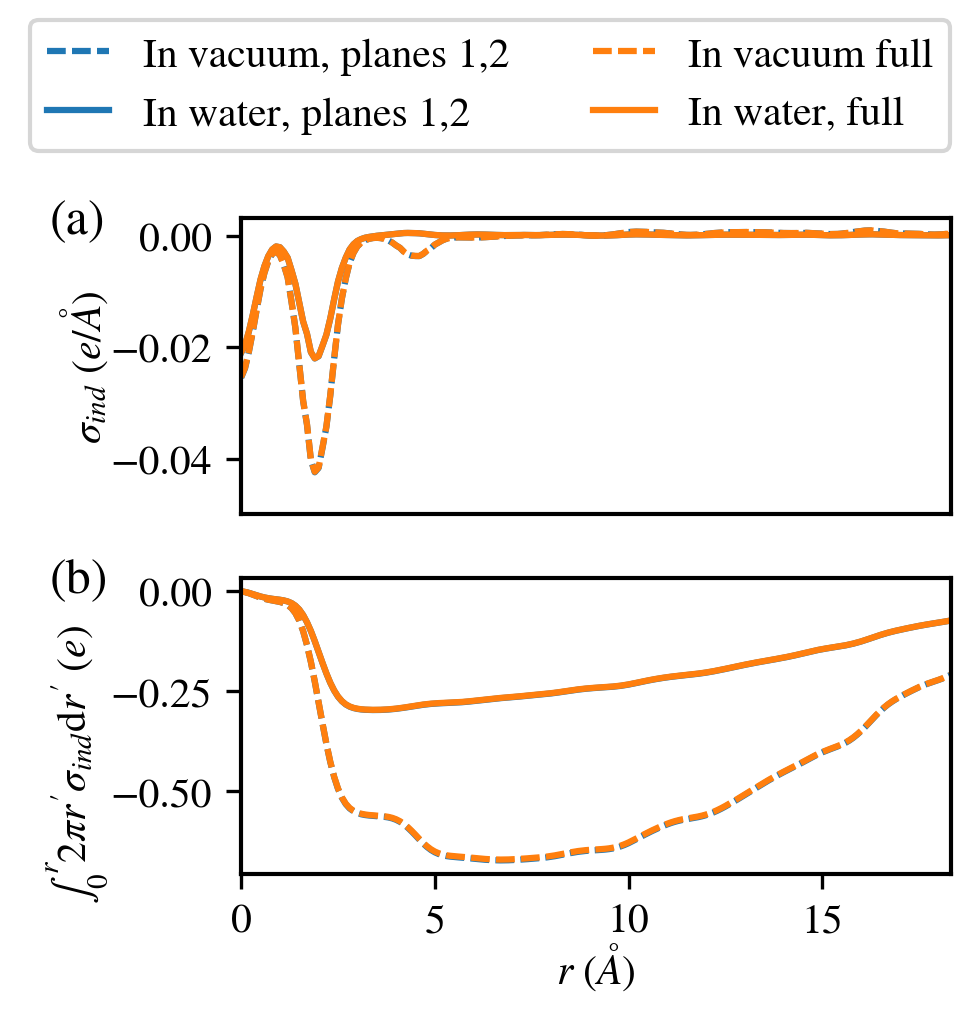}
  \caption{
  (a) Radial charge density profile, $\sigma_{ind}(r)$ and (b) Radial integral of $\sigma_{ind}(r)$, for a Na$^+$ ion in vacuum at various distances $z_{ion}$ from the first atomic plane of the electrode. In both panels, solid lines are the results for the system including the counterion (as in the main text), while darker dashed lines represent the results when the counterion is replaced by a uniform charge distribution among the atoms belonging to the first plane of the upper confining wall.}
\label{fig:CAvsUni}
\end{figure}

To test the influence of the localization of the counterion charge, we also performed calculations in vacuum where the counterion is replaced by a uniform charge distribution among the atoms belonging to the first plane of the upper confining wall. The comparison represented in Fig.~\ref{fig:CAvsUni} shows that this choice causes negligible differences in terms of induced charge profiles.

\section{Optimization of effective relative permittivity}\label{sec:relperm}

The effective permittivity $\epsilon_r^{eff}$ is determined for each distance $z_{ion}$ of the ion from the first atomic plane of the electrode by employing a simple parametric sweep optimization scheme. For each $z_{ion}$, we fit the radial integral of charge density profiles, $Q(r)=\int_0^r 2\pi r'\sigma_{ind}(r'){\rm d}r'$, obtained with molecular dynamics simulations by tuning the relative permittivity values in the continuum prediction Eq.~\ref{eq:continuum}. For this purpose, we introduce the following loss function: 
\begin{align}
    \mathcal{L}(\epsilon_r)= 
    \int_0^{L_x/2} \left[ Q^{MD}(r) - Q^{cont}_{\epsilon_r}(r)  \right]^2 {\rm d}r \, 
\end{align}
and define the effective permittivity as $\epsilon_r^{eff}=\argmin \mathcal{L}(\epsilon_r)$. Note that for the continuum prediction, we include the effect of periodic boundary conditions by summing over $41\times41$ periodic images to estimate the 2D density $\sigma_{ind}(x,y)$ before performing the radial average to compute $\sigma_{ind}(r)$ hence $Q(r)$. From Eq.~~\ref{eq:continuum}, it follows that this calculation can be performed once for $\epsilon_r=1$ and the radially averaged result scaled by $1/\epsilon_r$.

\bibliography{references}

\end{document}